\documentclass[a4paper]{article}

\usepackage{template/INTERSPEECH2021}

\title{Exploring emotional prototypes in a high dimensional TTS latent space}
\name{Pol van Rijn$^1$\thanks{Audio samples and supplementary figures can be found here: https://polvanrijn.github.io/gst-gsp/}, Silvan Mertes$^2$, Dominik Schiller$^2$, Peter M. C. Harrison$^1$, Pauline Larrouy-Maestri$^{1, 3}$, Elisabeth André$^2$, Nori Jacoby$^1$}
\address{
  $^1$Max-Planck-Institute for Empirical Aesthetics, Frankfurt, Germany\\
  $^2$Human-Centered Artificial Intelligence, Augsburg, Germany\\
  $^3$Max-Planck-NYU, Center for Language, Music, and Emotion, New York, USA}
\email{pol.van-rijn@ae.mpg.de, silvan.mertes@informatik.uni-augsburg.de, dominik.schiller@informatik.uni-augsburg.de}

\begin{document}

\maketitle

\begin{abstract}
Recent TTS systems are able to generate prosodically varied and realistic speech. However, it is unclear how this prosodic variation contributes to the perception of speakers' emotional states. Here we use the recent psychological paradigm `Gibbs Sampling with People' to search the prosodic latent space in a trained GST Tacotron model to explore prototypes of emotional prosody. Participants are recruited online and collectively manipulate the latent space of the generative speech model in a sequentially adaptive way so that the stimulus presented to one group of participants is determined by the response of the previous groups. We demonstrate that (1) particular regions of the model's latent space are reliably associated with particular emotions, (2) the resulting emotional prototypes are well-recognized by a separate group of human raters, and (3) these emotional prototypes can be effectively transferred to new sentences. Collectively, these experiments demonstrate a novel approach to the understanding of emotional speech by providing a tool to explore the relation between the latent space of generative models and human semantics.
\end{abstract}

\section{Introduction}
In the last decade, the quality of text-to-speech (TTS) has greatly been improved by the introduction of neural vocoders in combination with end-to-end TTS models like Tacotron \cite{shen2018, wang2017}. More recently, models have been proposed that are able to generate expressive speech, such as Tacotron with Global Style Tokens (GST Tacotron) \cite{wang2018}, Mellotron \cite{valle2020} and Flowtron \cite{valle2020a}. While these models can produce prosodically varied and realistic human-like speech, it is unclear how the prosody can be changed in a meaningful way such that it fulfills paralinguistic functions, like the communication of attitudes, intentions or emotions.

In order to find these prosodic representations, one needs to efficiently search the model's latent space. This is an increasingly difficult task for high dimensional spaces, since all combinations cannot be tried within reasonable time.
There are several psychological paradigms that can sample from such spaces using human participants, such as reverse correlation \cite{mangini2004}, Markov Chain Monte Carlo with People (MCMCP) \cite{sanborn2008} and Gibbs Sampling with People (GSP) \cite{harrison2020a}. GSP is a particularly recent paradigm that uses a continuous-sampling task instead of the binary choice task used by the other methods. This greatly increases information per trial and thus speeds up the parameter search.

Here we use GSP to search the prosodic latent space in a trained GST Tacotron model to explore prototypes of emotional prosody.

\section{Background}
There are two main challenges involved in synthesizing prototypes of emotional prosody. First, one must define a stimulus space, comprising of parametric manipulations applied to the sound. Second, one must find an effective way to identify regions of this stimulus space associated with particular emotional prototypes. 

One way to define the stimulus space is to construct a set of hand-crafted features that capture important aspects of prosody perception, for example pitch slope, jitter, and mean intensity. Previous work \cite{harrison2020a} showed that a simple handcrafted feature space was sufficient for generating distinctive, well-recognized prosodic prototypes of emotions. However, this approach is fundamentally limited because (i) it makes strong assumptions about which acoustic manipulations are relevant, (ii) not all potentially relevant manipulations to the sound can be made, because changing a single feature (e.g., pitch contour) regardless of other feature it is correlated with (e.g., spectral properties of the sound) may create unnatural and distorted speech, and (iii) when changing existing speech recordings, we are essentially changing continuous time-series, like pitch or intensity over time. Traditional hand-crafted features such as pitch slope and pitch range struggle to capture the full expressivity of underlying pitch or intensity contours.

Alternatively, the stimulus space may be created in a data-driven fashion. One solution is to use TTS models that factorize audio into separate text and prosody representations. GST Tacotron \cite{wang2018} is one of the most prominent examples of such TTS systems and is an extension of Tacotron, which is a sequence-to-sequence model learning the TTS task solely on pairs of recordings and transcripts (see supplementary Figure S1 for the architecture). In GST Tacotron a few components are added to Tacotron. A reference encoder \cite{skerry-ryan2018} is added, which compresses the Mel spectrogram to a fixed length embedding. This embedding is then passed to the so-called `style token layer'. This layer consists of a multihead attention mechanism, in which the given attention by the head is a similarity measure between the reference embedding and a bank of global style tokens. Based on the given attention weights a weighted average over all global style tokens is computed, which is called style embedding. Together with the text, the style embedding is passed to the Tacotron model which creates the predicted spectrogram. While this architecture can create varied speech, the control over prosody is relatively coarse, because the global style tokens are of a fixed length. Newer developments like Mellotron \cite{valle2020} and Flowtron \cite{valle2020a} aim to enhance prosodic control, which is a requirement for speech and song transfer. 

The second challenge is to identify regions of this space associated with particular emotional prototypes. A naive approach is to manipulate single dimensions independently, and assess the consequences for emotion perception. However, this assumes that the underlying dimensions contribute independently to emotion judgments, which cannot typically be justified. GSP provides a way to overcome this independence assumption: it leverages a well-established algorithm from computational statistics (Gibbs sampling) to identify regions of stimulus spaces associated with given semantic labels, while avoiding any independence assumptions \cite{harrison2020a}. 

\begin{figure}
    \centering
    \includegraphics[width=80mm]{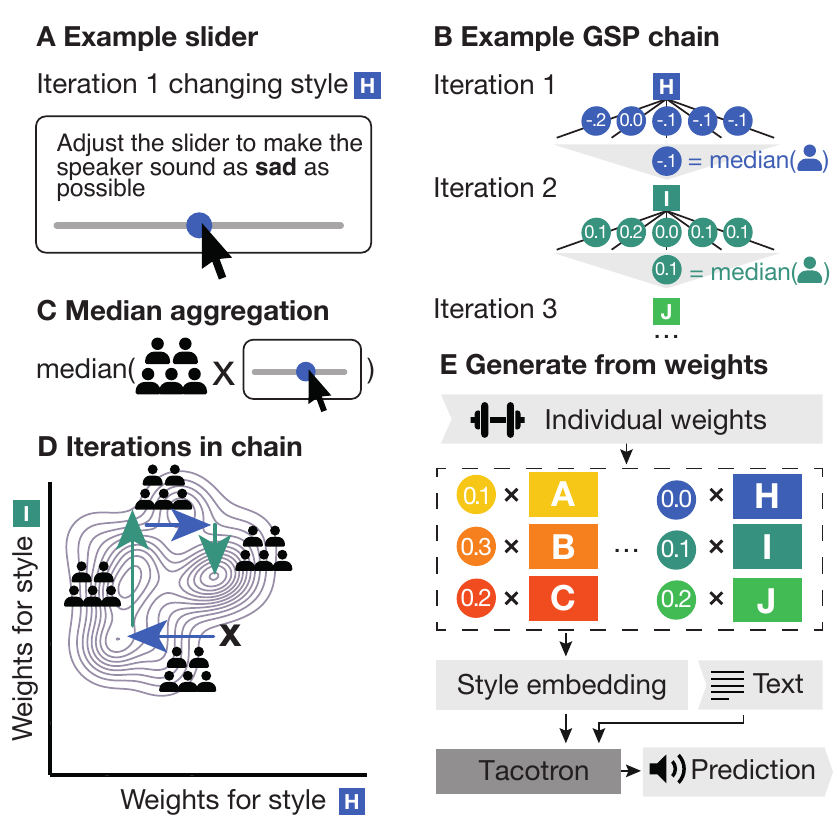}
    \caption{(A) Example slider in which a user is prompted to move the slider such that the speaker sounds as sad as possible. Moving the slider plays the sound with the selected attention weight (in this case for style H). (B) Schematic depiction of GSP chain. The chain consists of iterations. At every iteration only one dimension is changed. The colored dots represent choices by single participants. (C) Every slider is visited by 5 different participants. The median answer from all participants is passed to the next iteration, also compare with B. (E) Example for GSP process. For simplicity only showing dimension H and I. (E) GSP sliders control the attention weights that are passed to GST Tacotron that will create the stimulus for this configuration.}
    \label{fig1}
\end{figure}

\section{Methods}
\subsection{GSP}
GSP is an adaptive procedure whereby many participants collaborate to explore a high-dimensional sample space (Figure \ref{fig1}). The participants' responses are organized into sequences of iterations called ``chains'' (Figure \ref{fig1}B). A given iteration in a given chain has fixed values for all but one of the space's dimensions, and leaves the remaining dimension to be manipulated by the participants. In each trial, the participant is assigned to a particular iteration in a particular chain, and presented with a randomly initialized slider that manipulates the free dimension with real-time audio feedback. The participant is instructed to adjust the stimulus until it maximally resembles the target concept (e.g., \textit{sad}; Figure \ref{fig1}A). In our implementation, 5 different participants contribute trials for a given iteration in a given chain, and their responses are aggregated by taking the median (Figure \ref{fig1}C). This aggregated value is then propagated to the next iteration, where a different dimension is then manipulated. This procedure is repeated multiple times, cycling through each of the dimensions of the sample space (Figure \ref{fig1}D). The resulting process can be interpreted as a Gibbs sampler, a well-known algorithm from computational statistics for sampling from high-dimensional probability distributions \cite{harrison2020a}.

In the current experiment, participants change the attention weights of one of the 10 global style tokens.\footnote{We found that the attention weights of the four heads correlate with each other (average correlation of \textit{r} = .65). We therefore decided to reduce the dimensionality of the sample space by fixing each head to receive the same attention weight.} The participants are prompted to adjust a slider to make the speaker sound like a given emotion (see Figure \ref{fig1}A). The range of all dimensions is constrained to [-0.24, 0.38],  corresponding to a 94\% confidence interval of the attention weights given by the model in the training data, so as to minimize distortions. Every slider contains 32 equally-spaced slider positions. Since the synthesis of the stimuli must happen in real-time during the experiment, we used a Griffin-Lim vocoder for synthesis, finding that it achieved a good compromise between quality and speed (sound examples in the supplemental material). 
Every chain is initialized at 0 for every dimension, because extreme slider values can cause distortions to the signal.
\subsection{Synthesis model}
We train the model\footnote{https://github.com/syang1993/gst-tacotron} for 380,000 epochs using the same corpus (Blizzard Challenge 2013) and hyperparameters as the original paper \cite{wang2018}. When synthesizing from the model we set the attention weights (Figure \ref{fig1}E) directly from the current location of the relevant GSP chain in the sample space (Figure \ref{fig1}B), generating one output for each of the 32 possible slider positions. The participant would then select from these different outputs using the slider (Figure \ref{fig1}A).

\subsection{Material}
We use three phonologically balanced and semantically neutral sentences from the Harvard sentence text corpus \cite{harvardsentences}, and study three emotions: \textit{anger}, \textit{happiness} and \textit{sadness}. During the initialization of the experiment, a single sentence and emotion is assigned to every chain, such that every sentence and every emotion occurs equally often and are balanced across chains.
\begin{figure*}[h!]
    \centering
    \includegraphics[width=175mm]{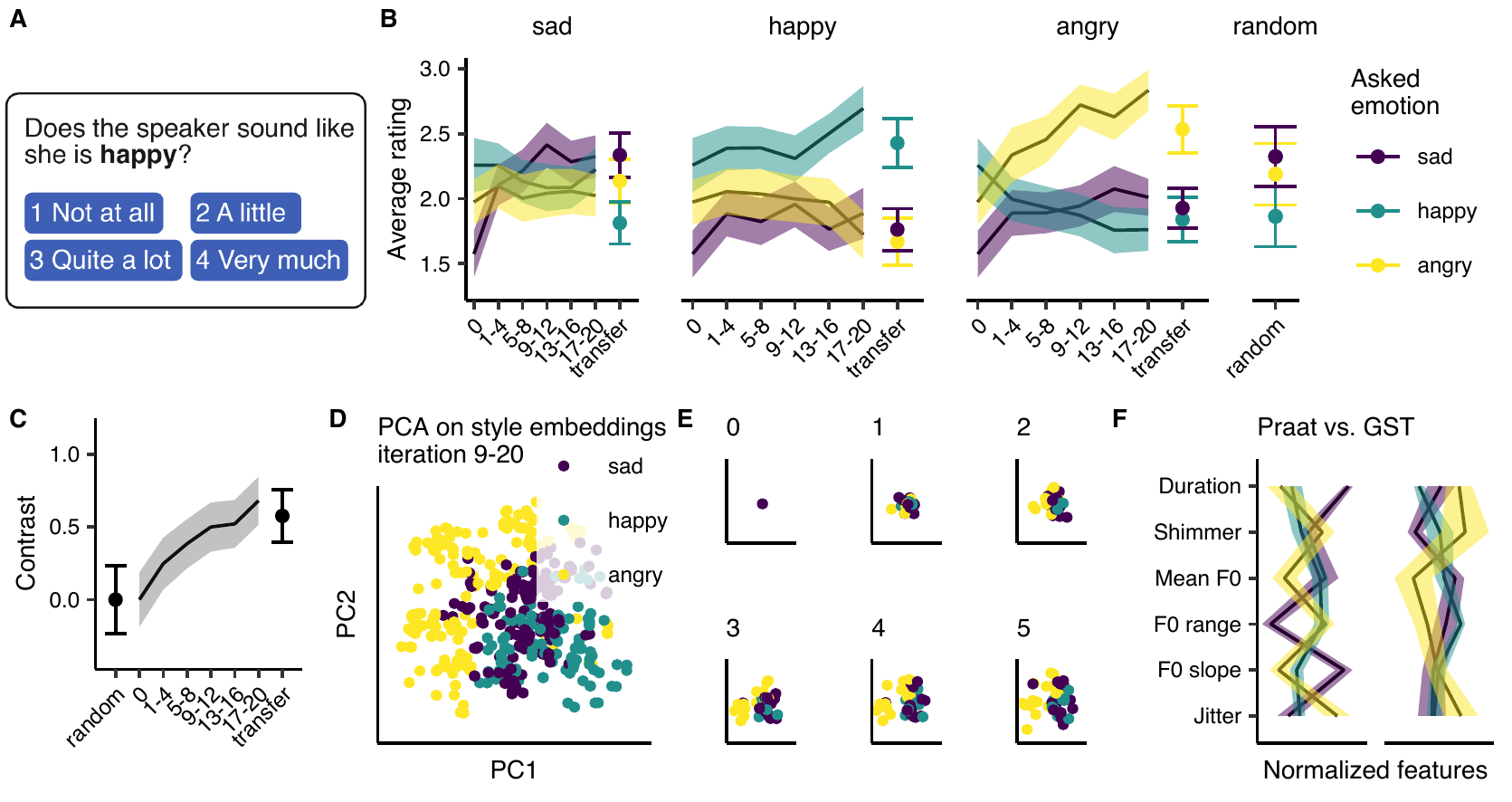}
    \caption{(A) Example validation trial. Audio plays automatically and user is prompted to answer. (B) Average ratings for the initial sample (iteration 0), binned iteration 1–4, 5–8, 9–12, 13–16, 17–20, the rating for the transfer and random samples (95\% confidence intervals). (C) Contrast between ratings (95\% confidence intervals).
    (D) Principal Component Analysis on style embeddings of 39 chains at iterations 9–20. (E) Development over iterations in PC style embedding space at iteration 0–5. (F) Comparison between the previous (changing specific acoustic features with Praat) and the current study (using GST Tacotron).}
    \label{fig2}
\end{figure*}

\subsection{Experiments}
130 US participants (61 female, 1 prefer not to say, 68 male) engaged in the experiment. The age ranged from 18 to 59 years old (\textit{M} = 36, \textit{SD} = 10). Before the experiment, the participant starts with three practice trials to get acquainted with the task. We terminated the experiment after 48 hours, after which 39 out of the 45 chains were full (20 iterations).

In a separate validation experiment, participants (\textit{N} = 82) rate how well samples matched each emotion on a four-point scale (see Figure \ref{fig2}A). The validation includes stimuli generated in the 39 full chains of the first experiment (i.e., the chains at different iterations of the experiment) as well as 18 random samples. We created 156 transfer stimuli by applying the median attention weights of the final GSP iteration to four novel sentences from the Harvard sentence corpus. These stimuli were also rated in the validation. On average every stimulus was rated 4.5 $\times$ for every emotion.

\subsection{Participants}
All participants were recruited from Amazon Mechanical Turk (AMT) and provided informed consent in accordance with the Max Planck Society Ethics Council approved protocol (application 2020\_05).  Participants were paid \$9/hour. Requirements for participation include a minimal age of 18 years, 95\% or higher approval rate on previous tasks on AMT (which helps to recruit reliable participants), residency in the US and wearing headphones (had  to pass a pre-screening headphone check \cite{woods2017headphone}).  Participant recruitment was managed by PsyNet \cite{harrison2020a}, an under-development framework for implementing complex experiment paradigms such as GSP and MCMCP. This framework builds on the Dallinger platform\footnote{https://github.com/Dallinger/Dallinger} for experiment hosting and deployment.

\subsection{Acoustic analysis}
In order to compare the current results with our findings from previous work \cite{harrison2020a}, we computed a similar set of acoustic features that were manipulated in the previous experiment. Duration- and pitch-related slider positions were well-recovered from the acoustical signal, were this was not the case
for the applied jitter- and tremolo-effect. We extracted duration, F\textsubscript{0} slope, mean and range\footnote{To make the range more robust to octave jumps, we do not use min-max range, but compute the standard deviation of the mean-centered pitch points.}, as well as shimmer (local) and jitter (ddp). All features were extracted with Praat \cite{praat} through a Python wrapper \cite{parselmouth}. To complement this hand-crafted feature set, we also computed a larger standard feature set (eGeMAPS) developed for detecting emotions from speech \cite{eyben2013, eyben2016}.

\section{Results and discussion}
\subsection{Validation}
As illustrated in Figure \ref{fig2}B, the ratings for the intended emotion steadily increase over the course of the iterations whereas the ratings for non-intended emotions plateau or drop. Moreover, there seem to be imbalances in the rating of the initial and the random samples, representing some perceived biases (e.g., iteration 0 sounds more happy than sad). To control for these imbalances, we compute the ``contrast'' between the ratings, corresponding to the mean rating for intended emotions minus the mean rating for not-intended emotions (Figure \ref{fig2}C). The contrast shows that the intended emotion reliably achieves higher ratings than the non-intended emotions. Consistent with the previous results, the contrast steadily increases over the iterations, but is close to 0 for the random samples and the initial sample (iteration 0). 

\subsection{Transfer}
Figures \ref{fig2}B and \ref{fig2}C also show average ratings for stimuli created by applying the derived attention weights to new sentences. These stimuli obtain high ratings, indicating that this transfer process worked remarkably well (see supplementary materials for audio examples).

\subsection{Structure of the latent space}
To investigate if the emotional sentences group together in the TTS latent space, we performed a Principal Component Analysis on all style embeddings of all stimuli in the experiment. Figure \ref{fig2}D depicts the first two principal components for the iteration 9–20. The figure shows that the three emotions separate moderately well on these two components. This grouping emerges relatively early on, providing additional support for early convergence of the GSP process (Figure \ref{fig2}E).

\subsection{Comparison to Harrison et al. (2020)}
In a previous study \cite{harrison2020a}, we used GSP to sample prototypes of emotional prosody when explicitly manipulating duration, loudness and pitch related features. Stimuli in the later iterations of the chains in both experiments are well-recognized as shown by validation experiments. However, Figure \ref{fig2}F shows that profiles computed on the stimuli from Harrison et al. (2020) look rather different from the profiles in the current study (\textit{r}(16) = .27, \textit{p} = \textit{ns}). Since these features only cover a very constrained acoustic space, we also compute the correlation between both studies on the larger feature set eGeMAPS. Again the correlation between both experiments is low (\textit{r}(256) = .31, \textit{p} $<$ .001), indicating that our GST experiment identifies different regions of the prosodic space than the experiment described in Harrison et al. (2020). To further address the question, we perform a 4-fold classification on both stimuli sets (linear kernel, C values: 1e-5, \dots 1e-1, 1). We include the last two iterations from Harrison et al. (2020) and iteration 9 – 20 from the current experiment to have a similar number of stimuli per experiment and made sure every emotion occurs equally often in every fold. We observed that the Unweighted Average Recall (UAR) is high for both experiments: Harrison et al. (2020) obtains 75.0\% UAR and the current experiment 79.4\% UAR (chance: 33.3\% UAR). However, when predicting Harrison et al. (2020) with the current results or vice versa, we obtain a lower UAR (49.1\% and 48.6\% respectively). These results suggests that both GSP methods generate samples with emotional states that occupy distinct parts of the feature space. However, there is only partial overlap between the features generated by the two methods. 

There are multiple potential explanations for this finding. First, the constrained feature set used in Harrison et al. (2020) might have forced participants to rely heavily on particular prosodic features that otherwise might be treated only as secondary emotional cues. Second, the two experiments rely on different speakers, and differences in their voices and accents may contribute to differences in emotional prototypes. Likewise, differences in the spoken texts may have contributed to differences in the resulting prototypes. These possibilities deserve further exploration.

\subsection{Limitations and outlook}
One clear limitation of the present study is that the prototypes might be stereotypical and might not fully represent how emotions are communicated in real life \cite{anikin2017a, barrett2019}. Future research could address this issue by replacing the discrete emotion labels in the GSP paradigm with descriptions of real-life emotional situations.

Future research could further improve the parametrization of the latent space, for example by relaxing the constraint that each head receives the same attention weight, using different TTS models, and training on different datasets. Most importantly, future research will need to test more heterogeneous populations and also train on non-western and non-English corpora in order to make valid claims about emotional prosody and to develop robust applications \cite{henrich2010}.

\section{Conclusion}
In this paper, we used Gibbs Sampling with People together with a trained GST Tacotron model in order to explore prototypes of emotional prosody. Our results show that (1) particular regions of the model's latent space are reliably associated with particular emotions, (2) the emotional prototypes are well-recognized by human raters, and (3) the emotional prototypes can be transferred to new sentences. We showed that the emotional prototypes occupy different positions in the TTS latent space and do so from early stages of the experiment, indicating early convergence. Finally, we found interesting acoustic differences between the current study and Harrison et al. (2020), which should be explored in future research by carefully comparing emotional prototypes created with hand-crafted acoustic manipulations versus those created by TTS models. All in all, GSP in combination with GST Tacotron seems to be a useful and efficient tool for studying emotional prototypes, for exploring speaking styles in existing TTS systems, and for generating new emotional sentences based on pre-existing speech recordings.

\section{Acknowledgments}
This work has partially been funded by the European Union Horizon 2020 research and innovation programme, grant agreement 856879.

\bibliographystyle{IEEEtran}

\bibliography{bib}{}

\end{document}